Draenert G.F.*†,  Mitov G.**

# Lack of corundum, carbon residues and revealing gaps on dental implants

* Prof.Dr.Dr.Dr. Guy F. Draenert, MD DMD PhD, University of Marburg and Implant Institute, Tal 4, 80331 Munich, Germany

guydraenert@gmail.com

guy.draenert@staff.uni-marburg.de

** Prof.Dr. Gergo Mitov, DMD PhD, Clinic for Prosthetics, Danube Private University, Krems, Austria

†       Corresponding author

GRAPHICAL ABSTRACT

# Lack of corundum, carbon residues and gaps on dental implants

nearly corundum-free surface

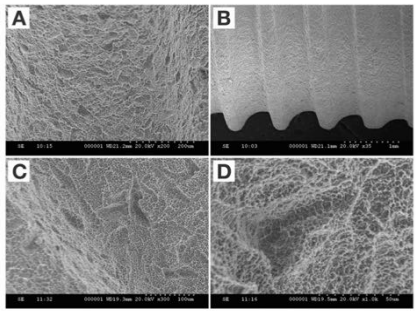

gaps indicate disappeared mater

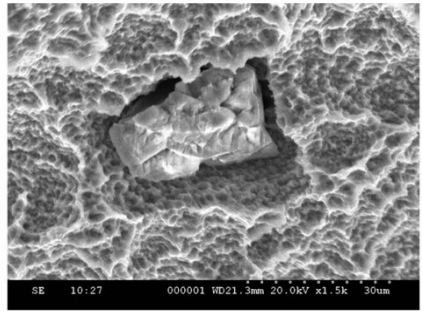

unclear dirt with high carbon levels

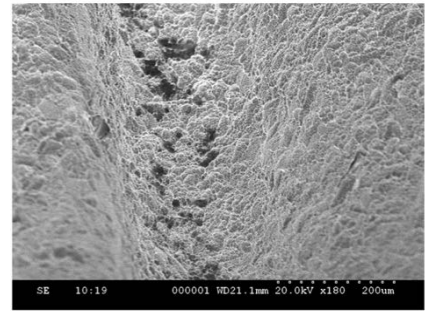

patented dextrane protection technology: WO 002018/189185

# ABSTRACT

Surface modification is an important topic to improve dental implants. Corundum residues, which are part of current dental implant blasting, disappeared on Straumann dental implants in recent publications. In our investigations of the surface of 4 different Straumann implants using scanning electron microscopy (SEM) and energy-dispersive X-ray spectroscopy (EDX) we found the following three main findings: surfaces are nearly corundum-free, disseminated gap-framed corundum particles and significant molecular carbon residues. The data strongly suggest that Straumann applies a modified surface technology on dental implants to remove corundum residues and involving unclear carbons. One explanation could be, a Straumann patent involving a dextran coating allowing easy corundum particle removal by aqueous solution, while unintended molecular carbon residues cannot explain all findings. This change of the production process without a new approval by the FDA would be a violation of US federal law and the carbon bindings are a possible danger to patients.

# INTRODUCTION

The industry In general has often tried to outsmart federal law to solve their economically driven problems arising from regulatory requirements as recently done e.g. by Volkswagen with the diesel fraud [1]. The biggest danger concerning invasive medical devices in this concern is the field of long-term complications as known from hip and breast implant scandals [2, 3].

The field of dental implants is a huge market. The worldwide implant dentistry market is estimates at $US 4.2 bn (1 CHF/1.12 $US). Straumann Inc. is a major dental implant player and defines the USA as its major target market with an estimated 2.5 mio dental implants placed each year [4]. Straumann itself estimates their own market share with 24% resulting in $US 1 bn with a CAGR of minimum 4% on the basis of 2017 [5]. Approvals for new surfaces of dental implants, which are invasive medical devices (class II), with the FDA and other authorities require expensive clinical trials. It is therefore an important option to claim equality status to an already approved device. This limits the costs and loss of time for clinical trials to a minimum [6]. Straumann did so, when applying for 510k premarket notification and approval of the SLActive surface enosseous dental implants by the FDA in 2006 [4]. Therefore, both current surfaces SLA and SLActive claim the same FDA approval.

Dental implants become osseointegrated in the jaw bone by osseous healing of the bone onto the implant surface. The basic principles of bone and biomaterials are described elsewhere [7-9]. The osseointegration of dental implants involves bone healing with its four phases: aseptic inflammation, soft callus, hard callus and remodelling. The primary mechanical stability of the implant in the jaw bone is thereby declining whereas the secondary stability by direct mineralized bone tissue connection to the implant analogous to the hard callus, is rising. This process takes 3-6 months after implant insertion.

The implant surface plays an important role in this process. Rough surface promotes the osseointegration. Current techniques involve a two-step surface treatment as standard for most titanium implants. The structuring involves blasting with either corundum or other materials for macro structuring and etching with fluoric acid to result in a rough microstructure. The measurement of this surface microstructure and the biological value concerning osseointegration are subject to complex discussions [10]. Rupp et al. also describe several aspects of tertiary surface treatments to improve microstructure, antibacterial properties and hydrophylisation, the important aspects are nevertheless blasting and etching [10]. This publication introduced the unclear and unproven term "bio-carbon" in association with Straumann implants. In the early days of dental implantology smooth machined implant surfaces were applied and do still perform well in the ongoing long-term studies. However, animal experiments showed a clear benefit for blasted and etched implants as shown in several mile-stone studies and became gold standard as primary and secondary surface treatment in dental implant production [11-13]. Sandblasting is usually performed with either titanium oxide or aluminum oxide (corundum) particles. Etching is done in fluoric acid strong enough to attack the stable titanium oxide inhibition layer on the titanium surface [14]. Corundum residues crushed into the soft titanium metal are an inevitable side effect of the blasting [15-18]. These residues are therefore an accepted status by the FDA.

The described sandblasting and etching technologies are applied in both current Straumann surfaces SLA and SLActive [14, 17, 19].

In the following years it became a trend to replace corundum with titanium dioxide for blasting dental implants. The main driver for this development was a rising discussion about corundum particle residues and possible disadvantages of corundum [17]. Particles of titanium oxide are difficult to prove on the surface and avoids the discussions without being more clean. An animal study showed clearly no disadvantages of corundum residues on the osseous healing of implants in biomechanical and histomorphological examination [20]. However, almost all premium manufacturers changed the production process to titanium dioxide blasting, while Straumann kept to corundum and was facing further ongoing critics by industry competitors due to its blasting procedure and associated particle residues [18].

Recently published data show Straumann implants completely free of any corundum residues [21]. The physical impact of corundum particles makes residual particles on the titanium surface inevitable. This scientific paradox was the basis for the presented study. The primary aim of this study was to investigate corundum residues on the surface of Straumann implants using scanning electron microscopy and energy dispersive x-ray (EDX) analysis. The secondary aim of the study was to examine aluminum and other residues on the surface of Straumann SLA and SLActive implants and in the storage liquid of SLActive implants.

MATERIALS AND METHODS

**Dental implants**

We used different Straumann dental implants (n=4) for this study es decribed below. SLA implants are sandblasted and etched, whereas SLActive implants are additionally hydrophilized as described above. The information concerning the production date and site were provided by Institut Straumann AG. These samples are:

- Sample #1 (SLActive): LOT RA939; REF021.2308; Straumann BL SLActive Titanium 3.3mm / 8mm; volume of storage solution: 2.818 ml; distributor: Straumann Germany GmbH; production date: 27th april 2018; production site: CH / Villeret.

- Sample #2 (SLA): LOT PX142; REF021.3412; Straumann BLT SLA Titanium 3.3mm / 12mm; storage: dry; distributor: Straumann Germany GmbH; production date: 2nd march 2018; production site: CH / Villeret.

- Sample #3 (SLActive): LOT NX064, REF033.652S; Straumann Standard Plus SLActive Titanium 4.8mm / 10mm; volume of storage solution: 2.875 ml; production date: 10th august 2017; production site: CH / Villeret.

- Sample #4 (SLActive): LOT KY861, REF0033.762S; Straumann Tapered Effect SLActive Titanium 4.1mm / 10mm; volume of storage solution: 2.873 ml; production date: 12th february 2016; production site: CH / Villeret.

The samples were opened and processed under clean conditions in the SEM laboratory (University of Dusseldorf, Department for Oral Surgery) only. We performed two examination sessions in august and september 2020 to allow appropriate time for consideration of primary results before the secondary examination. Samples were stored in sterile vials and air closed between the sessions.

**Imaging and EDX (energy dispensive x-ray analysis)**

We used a Hitachi S-3000N scanning electron microscope with an EDAX Detection unit PV7746/33 ME (University of Duesseldorf, Department of Oral Surgery Prof Becker). The devices were maintained correctly, calibrated and the EDAX was freshly fueled with nitrogen all in compliance with the instruction manuals. The samples were placed on the SEM carrier module under clean conditions. The technician was trained on the devices. The specific settings are mentioned with the results below. Samples were used native without spattering. The measurements were used as descriptive support of the morphological findings in this combined SEM-morphological/EDX examination. The implants surfaces were first examined at a low magnification of 200x scanning over the whole implant surface that was accessible in the given position in the SEM device, which is approximately 60-70% of the surface and higher magnification (up to 1000x) was used when visual anomalies were detected as described below. We checked the correct surface structure (sandblasted and etched) and looked for residues of any kind, which were then examined in more detail at higher magnification and EDX.

**SEM and EDX examination of dental implants**

The packages of the dental implants were opened under clean conditions and exclusion of exterior contamination in the SEM examination room of the University of Düsseldorf, department of oral surgery. The liquid of the SLActive implants was harvested with a Gilson Pipetman 200 in sterile Epicaps. The implants were vacuum dried for the first examination and rinsed with pharma-grade, sterile demineralized water 3 times and vacuum-dried also for the second examination. The presented EDX and SEM examinations refer to the samples as mentioned below in detail.

**EDX examination of SLActive fluid**

We applied the EDX in order to qualitatively analyze the storage liquid of the implants. SLActive implants are stored in a liquid in their vials and not dry. Any residues on the implant surface are subject to mixing into the this liquid. We applied EDX examination to evaluate this liquid in order to solve the problems associated with other methods. All other methods required larger volumes or more complex procedures providing quantitative data also. The desired method of choice would have been mass spectrometry of the storage liquid. However, the necessary volume required by certified environmental analysis institutes in Germany are 5 ml or more, exceeding the given storage liquid volumes as mentioned above. We therefore applied an electron-reflecting carrier (gold) and placed microliter volumes to apply the EDX-device for this analysis. An United States Gold Eagle 1/10 oz. coin was cleaned with acetone and placed on the SEM carrier module under clean conditions. 20 $\mu$l Storage liquid were applied at the designated sample positions as described below with a Gilson pipetman 200. The samples were dried in a vacuum chamber as usual and then transferred in the SEM device.

RESULTS

**EDX examination of SLActive fluid**

The EDX examination of the samples as shown in fig. 2 and 3 showed the following difference between the empty control run and the samples:

Sample #1 (LOT RA939): C (carbon) 50.25 Wt%/ 70.28 At%; Na (sodium) 23.97 Wt%/ 17.51 At%; Cl (chloride) 25.77 Wt%/ 12.21 At%.

Sample #2: SLA has no storage fluid.

Sample #3 (LOT NX064): C (carbon) none; Na (sodium) 8.67 Wt%/ 35.28 At%; Cl (chloride) 9.73 Wt%/ 25.67.

Sample #4 (LOT KY861): C (carbon) none; Na (sodium) 4.15 Wt%/ 16.35 At%; Cl (chloride) 3.70 Wt%/ 9.45 At%. Values below 1 Wt% or At% are excluded and considered not relevant. There was no aluminum whatsoever in any speciemen, not even in percentages below 1%.

**EDX-findings on the intraosseous-dedicated rough surface areas**

We performed two generally different examination sessions as described above. EDX was applied in a random sampling pattern. The first session (S1) was done with SLActive solution dried on the surface. Session 2 was performed after rinsing with water (S2).

Carbon (C) can be vastly found in all areas with and without the association of particles as describe below on all samples. Values in the free areas vary (Wt% describing the mass weight percentage relative to the elements described, while At% representing the atomic percentage relative to the elements describe). These are the surface carbon data without association to particles:

Sample #1 (LOT RA939): **S1:** 5.34 Wt%/ 32.13 At%;36.37 Wt%/ 68.57 At%; 16.92 Wt%/ 32.98 At%; 20.68 Wt%/ 38.98 At%; 23.35 Wt%/ 47.47 At%; 20.10 Wt%/ 44.82 At%; 20.27 Wt%/ 44.38 At%; **S2:** none

Sample #2: (LOT PX142): **S1:** 65.78 Wt%/ 79.82 At%; **S2:** 7.19 Wt%/ 20.56 At%; 6.91 Wt%/ 19.07 At%;

Sample #3 (LOT NX064): **S1:** 3.39 Wt%/ 9.56 At%; **S2:** 10.25 Wt%/ 24.67 At%; 10.17 Wt%/ 24.45 At%;

Sample #4 (LOT KY861): none;

**Particle and blasting examination**

The presented EDX and SEM examinations of particles refer to the rinsed samples only.

*SEM imaging of corundum blasted particle-free areas*

Corundum blasting results in a specific surface macrostructure (see fig.4). However the typical and physics-based corundum residues, that are struck in the titanium surface with a sharp and gap-free interface to the titanium, can only be found rarely (see fig.5). Unusual cave-shaped defects with high carbon surface values of 25.65 Wt%/ 50.49 At% can be found also (see fig.6).

*SEM imaging and EDX (energy dispersive x-ray analysis) of unusual particles*

Particles of various shape with aluminum and carbon loading can be found. All of these particles show a gap at the interface between the particle and the titanium surface. Some of these particles show a corundium-typical structure (see fig.7). Some particles are more smooth morphologically (see fig.8).

DISCUSSION

The most important three findings of this study are as follows. A nearly corundum-free surface shows a typical surface structure of corundum sandblasting and etching. Disseminated corundum particles show coronal gaps indicating disappeared material of any kind. The surface shows extended molecular carbon residues. We will discuss these three main findings in detail below. The crucial point is to understand that titanium does not expand and corundum does not shrink under the conditions applied in Straumann implant production and the examinations performed in this study. The gaps shown in fig.7 and fig.8 are implicating a material removal during production as discussed in detail below. This change in the production process explains all thre main finding.

***EDX examination and SEM methodology***

It is important to emphasize that EDX is a method to detect molecules, whereas particle residues are detected in SEM morphologically and further analyzed in EDX examination on a molecular level. Scanning electron microscopy (SEM) as morphological method was combined with energy dispersive x-ray (EDX) analysis to reveal the presence of particles molecular elements on the surface of the specimens. It is important to understand the definition of Wt% describing the mass weight percentage relative to the elements described, while At% representing the atomic percentage relative to the elements described.

*Confusing publications concerning Straumann implant surfaces*

The Straumann SLActive surface is a hydrophilized surface additionally to blasting and etching compared to the SLA surface which is blasting and etching only [22]. They claim that a treatment with chloric and sulfuric acid leads to the hydrophylisation by removing unclear surface hydrocarbon residues and changed surface free energy resulting in higher hydrophylisation when applying oxygen reduced conditions by nitrogen gas and storage in a physilogical sodium chloride saline. The FDA approved the implant with "equality status" to the previous SLA surface based on this claim as described above. Following papers did not specify the technology either [6]. The authors also claim that unclear biocarbonates become removed by the SLActive process and result in better hydrophylic properties. However there is no examination of the concrete molecular nature of these carbonates and their origin neither in this publication nor in any other. These carbon residues are the main concern for the safety and health of patients and it is completely unclear where they come from or if they are even actively applied in any production process. Straumann remained very active in the field of dental implant surface technologies [23-28].

Recent papers showed controversial results concerning the corundum residues on SLA and SLActive surfaces indicating any possible technology application by the company. A swiss study group showed vast corundum residues on SLActive surfaces in their paper [18]. Older papers showed similar corundum residues after blasting on the surface of Straumann and other implants too [15-17]. Duddeck et al. claimed no corundum particles on Straumann SLA surfaces and described the Straumann implant is "clean on molecular level" [21].

We found vast molecular carbon residues of unclear nature and origin and rare but never the less present corundum particles on SLA and SLActive surfaces. This proves that Straumann dental implants are not clean on a molecular level. Those unclear carbon bindings can be a serious threat unless ruled out by in-depth examination and proven harmless.

However, avoiding titanium blasting can be an advantage too. Titanium oxide is intensively discussed to be highly problematic [29-33]. The Straumann blasting technology could therefore be one of the best on the market except of the unclear carbon threat.

### *SLActive solution*

We cannot find any trace of free molecular aluminum in the SLActive storage solution. However, it was possible to detect high carbon values in one of the samples as described above.

### **Carbon and gaps**

The main findings are molecular carbon residues on the titanium surface and unclear gaps between titanium surface and the disseminated corundum particles. There are several explanations possible for these observations. Yet combining them leads to limited options and a conclusion as discussed below. Unintended carbon residues include plastic as often described in the literature. Plastic devices during production and packaging are discussed and associated plastic contamination has been analyzed by confocal microscopy [34]. Intended carbon residues are another option leading to multiple explanations that fit to desired corundum removal involving the gaps observed in our results.

*Unintended carbon dirt from the FDA-approval-conform production process and optimized corundum blasting*

There are few publications dealing with dirt on unused new dental implants. Carbon residues in general are described even more rarely. Straumann created the common opinion that carbon residues are regular residues of the production process. This involves in the SLActive key paper also [22]. The authors mention "bio-carbonates" repetitively in this publication. It appears that this mysterious material is the planned excuse for carbon on implants. They also state that corundum residues of 16-18% after blasting are a normal value. The citation behind this number is a mongraphy section written by Wieland M. who was part of the development team of the SLActive surface at Straumann company [35]. While Rupp et al. mentioned aluminum oxide residues in the SLActive paper, they did not specify the relation to the SLActive technology and the precise technology behind the acid application and nitrogen or fluid storage [22]. We doubt that the described amount of carbon residues can be explained with rare usual contamination during production anyhow.

Corundum blasting can be optimized in terms of blasting angle, particle size and speed. However it appears highly unlikely that any optimization can result in almost zero corundum residues. It also does not explain the observed gaps.

*Intended surface technologies leading to carbon in EDX examination*

The most appropriate explanation for the molecular carbon residues found on Straumann implants in our study is any carbon compound that is part of a production process technology and remained thereafter. This includes plastics as well as sugar or other carbon-containing materials. This technology can be a corundum removal technology. This is realistic, considering the market pressure on Straumann to solve its corundum particle problem.

*Corundum removal technology*

Three possible explanations fit to the results and enable easy corundum removal in terms of an intended production technology. This includes: i.) blasting with corundum and separate carbon compound particles, ii.) carbon compound coated corundum particles or iii.) carbon compound coating of the implants before corundum blasting. The fluoric acid etching after blasting can lead to uncalculated carbon residues as an unoptimized production process. This could explain the high carbon surface levels.

*Gaps at the particle interface*

Regular corundum (aluminum oxide) blasting results in particle residues on the titanium implant surface as cited above [15-18]. Straumann SLA and SLActive implants are processed the same way concerning blasting as claimed in the related FDA 510k application [4]. Straumann implants were shown to contain these regular residues in several publications as mentioned above. [17, 18]

Solid particles do not shrink and titanium surface defects do not expand after blasting and under the processing and examinations done in this study. Therefore, the images fig. 7 and 8 are utmost important to understand the conclusion of the results. The only explanation that is sound concerning these morphological results is that there was a material present during the blasting process between titanium surface and particle that was removed later.

*Straumann implant surface research and development*

An active research and development group within Straumann includes Simon Berner as staff. He participates in several patent publications that deal with surface treatment modifications [23, 24, 26-28]. These technologies include the application of sugar, fibrin, titanium nano structures and phosphates. The most important technology related to our results is the patent application WO2018/189185 / US2020/0078142 dealing with a dextran coating on dental implants [27, 28]. The method fits perfect to our observations and solves the current problem of Straumann concerning corundum blasting residues. The coating does allow to remove the blasting particles easily by rinsing with water. The invention was filled in 2018 which also fits to the production date of our samples and the communicated production change as suggested by Duddeck D in a personal communication. There are no publications of Simon Berner and the dextran technology in particular. The detail that this patent was filed for ceramic implants can be an intended mislead as well as the unsatisfying explanations for carbon residues in Straumann originated publications so far.

Changing the production process is subject to a new FDA-approval. The unclear threat of unidentified carbon residues of high surface percentage is to be examined in-depth.

# CONCLUSION

The data strongly suggest that Straumann applies a modified and non-approved surface technology involving carbon compounds on dental implants to remove corundum residues. This change of the production process without a new approval by the FDA is a violation of US federal law and a possible danger to patients.

# ACKNOWLEDGEMENTS

We thank Mrs. Tina Hagena and the Department for Oral Surgery, University of Dusseldorf for their material support in scanning electron microscopy by gaining access to their SEM and EDX unit. We thank Dr. Nilton Penha, Av. Rio Branco, 26, Sobreloja – Centro, Rio de Janeiro, Brasil for the communication and the hint to Straumann implant surface patents with sugar. We thank Prof. Thomas Tischer from the University of Rostock, Germany for contributing to the manuscript.

# DECLARATIONS

This study is not a clinical trial. Ethics approval and consent to participate are not applicable. Availability of data and materials is possible at any time by contacting the corresponding author. The authors declare that they did not receive any financial benefits of any kind of any company or other player concerning this publication at any time or any other concern during the last 24 months. They are not related to any company or get payments or benefits orf other kind of any company during the last 24 months.

# Captions (figure legends)

Fig.1:

The coin model used to allow EDX analysis of the storage liquid in the EDX. The 3 positions of the 3 SLActive samples are: #1: Sample #1 (LOT RA939); #3: Sample #3 (LOT NX064); #4: Sample #4 (LOT KY861).

Fig.2:

Energy dispersive x-ray (EDX) analysis without SLActive storage solution as described in detail in the results. The sample positions refer to the described letters in "Unite States of America" on the 1/10 oz. golden eagle coin.

Fig.3:

Energy dispersive x-ray (EDX) analysis with SLActive storage solution as described in detail in the results. The sample positions refer to the described letters in "Unite States of America" on the 1/10 oz. golden eagle coin.

Fig.4: Predominantly morphologically clean surface with typical corundum blasting pattern including disseminated deep bumps and almost no corundum residues: A: sample #1 (200x); B: sample #1 (35x);  C: sample #2 (300x); D: sample #2 (1000x).

Fig.5: Typical corundum particle struck in the titanium surface without gap (sample #1; 800x).

Fig.6: Unusual surface excavations with high carbon levels (sample #1; 180x).

Fig.7: Typical corundum particle struck in the titanium surface with a gap indicating the removal of any material that was located there at the time of blasting (sample #1; 800x).

Fig.8: Untypical particle with aluminum content indicating corundum struck in the titanium surface with a gap indicating the removal of any material that was located there at the time of blasting (sample #4; 600x).

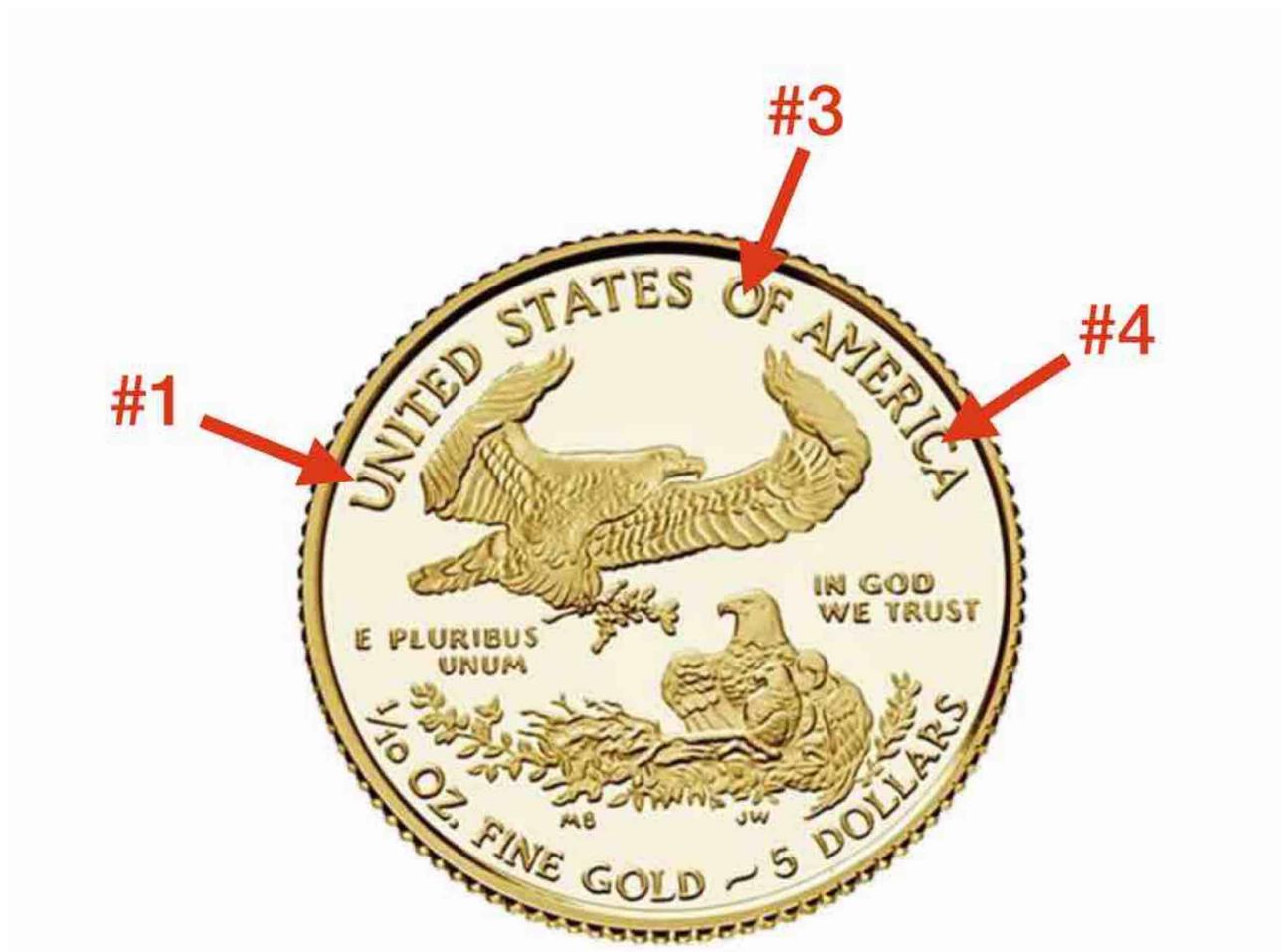

Fig.1:

The coin model used to allow EDX analysis of the storage liquid in the EDX. The 3 positions of the 3 SLActive samples are: #1: Sample #1 (LOT RA939); #3: Sample #3 (LOT NX064); #4: Sample #4 (LOT KY861).

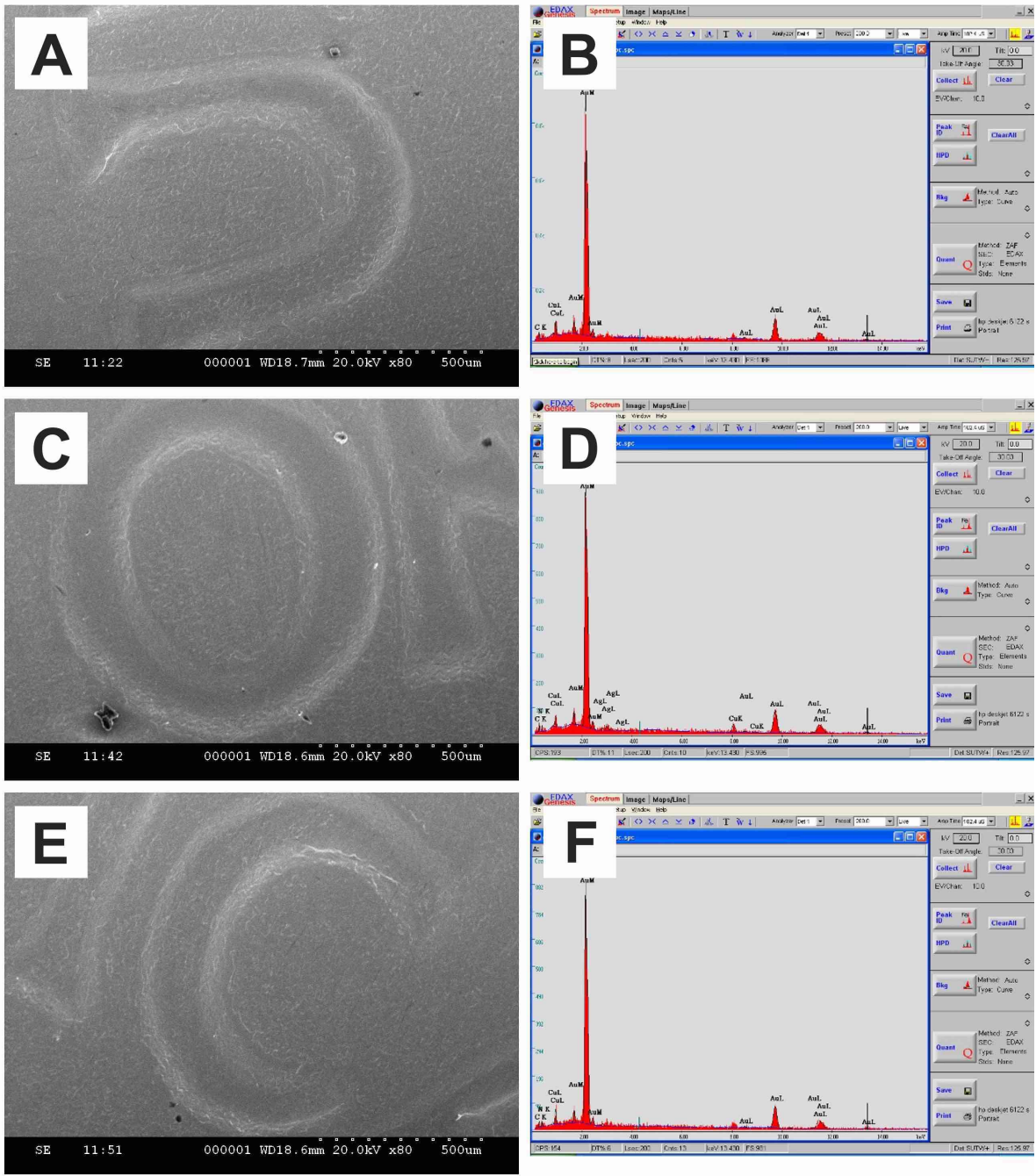

Fig.2:

Energy dispersive x-ray (EDX) analysis without SLActive storage solution as described in detail in the results. The sample positions refer to the described letters in "Unite States of America" on the 1/10 oz. golden eagle coin.

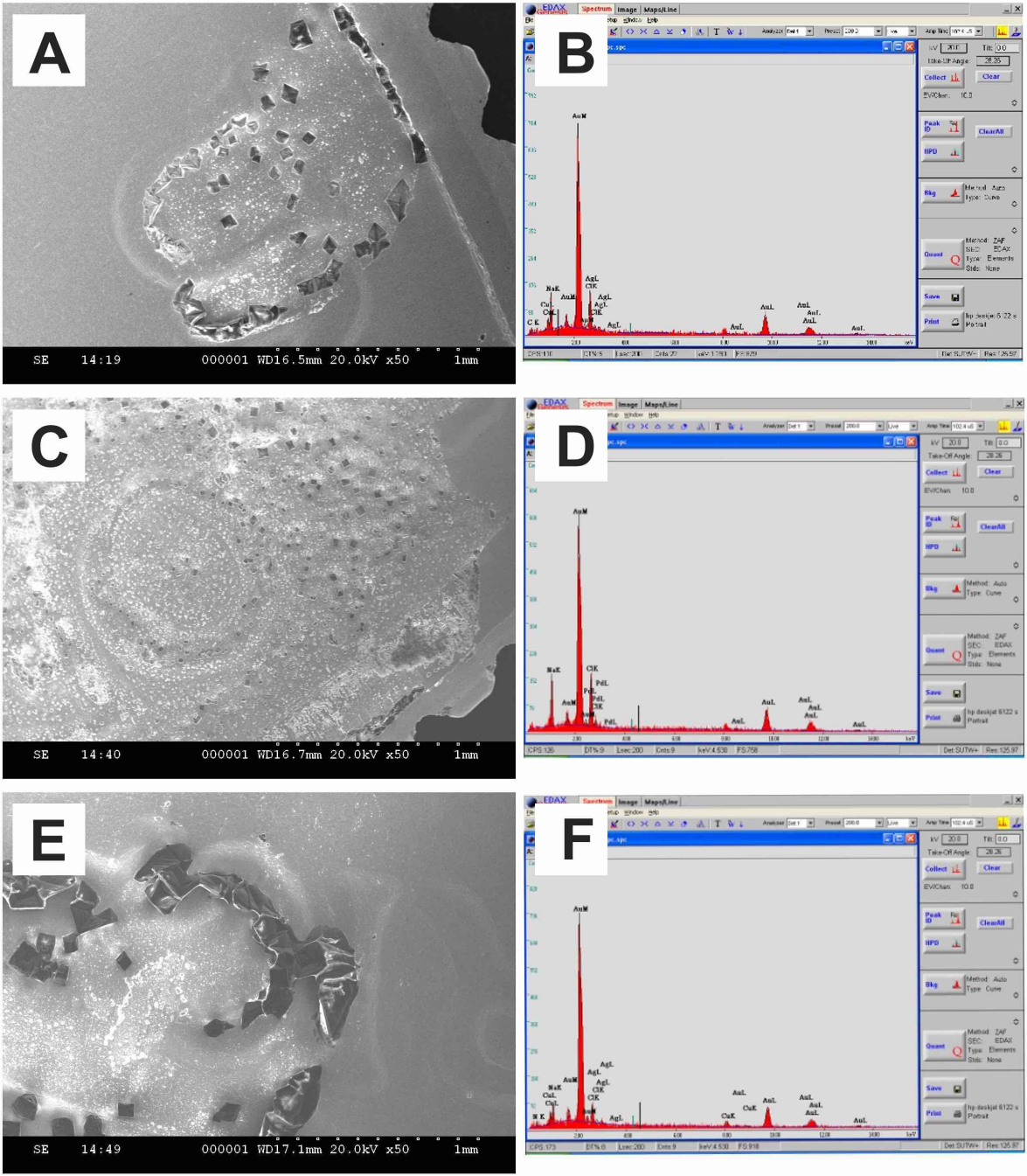

Fig.3:

Energy dispersive x-ray (EDX) analysis with SLActive storage solution as described in detail in the results. The sample positions refer to the described letters in "Unite States of America" on the 1/10 oz. golden eagle coin.

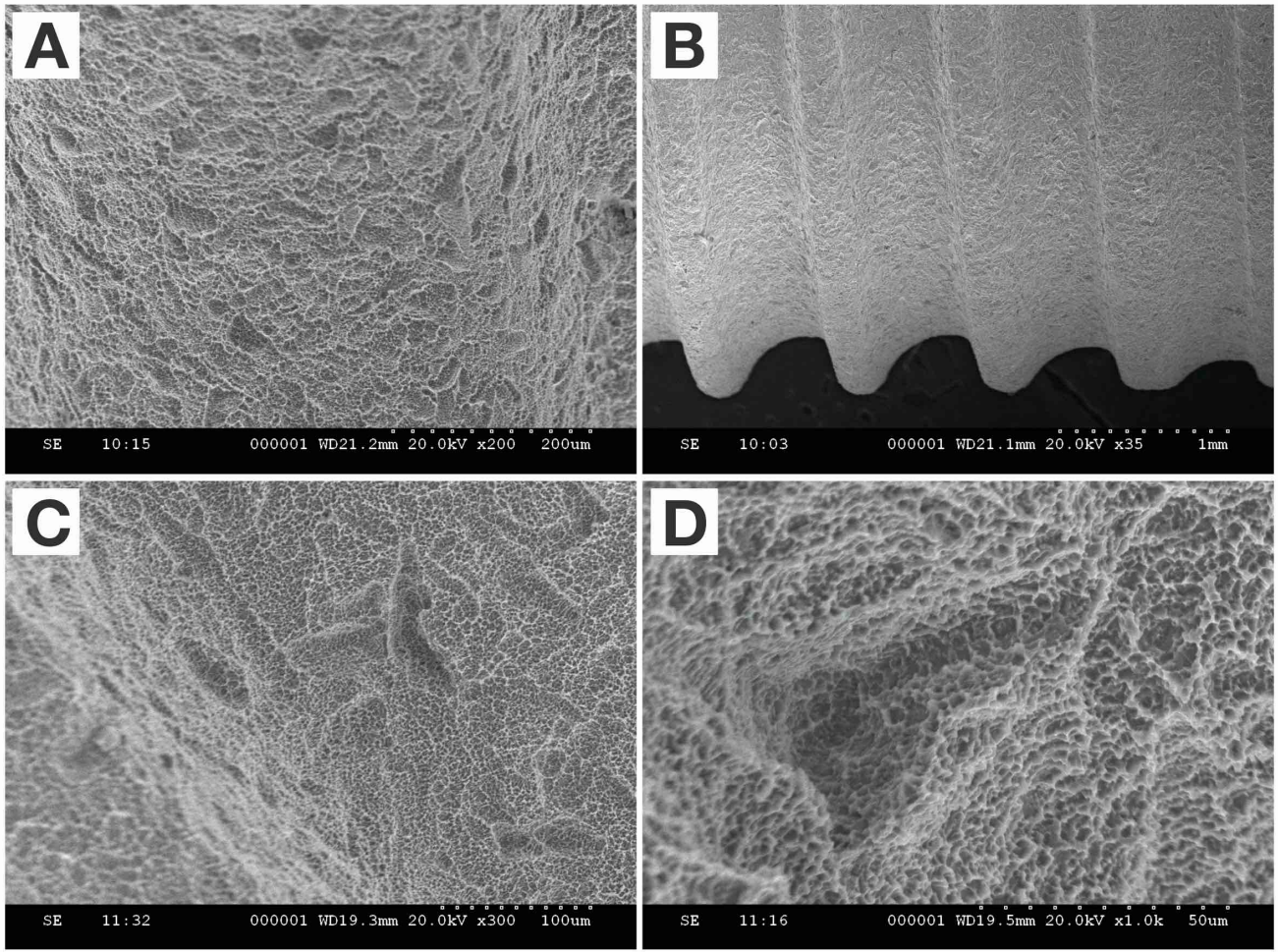

Fig.4: Predominantly morphologically clean surface with typical corundum blasting pattern including disseminated deep bumps and almost no corundum residues: A: sample #1 (200x); B: sample #1 (35x); C: sample #2 (300x); D: sample #2 (1000x).

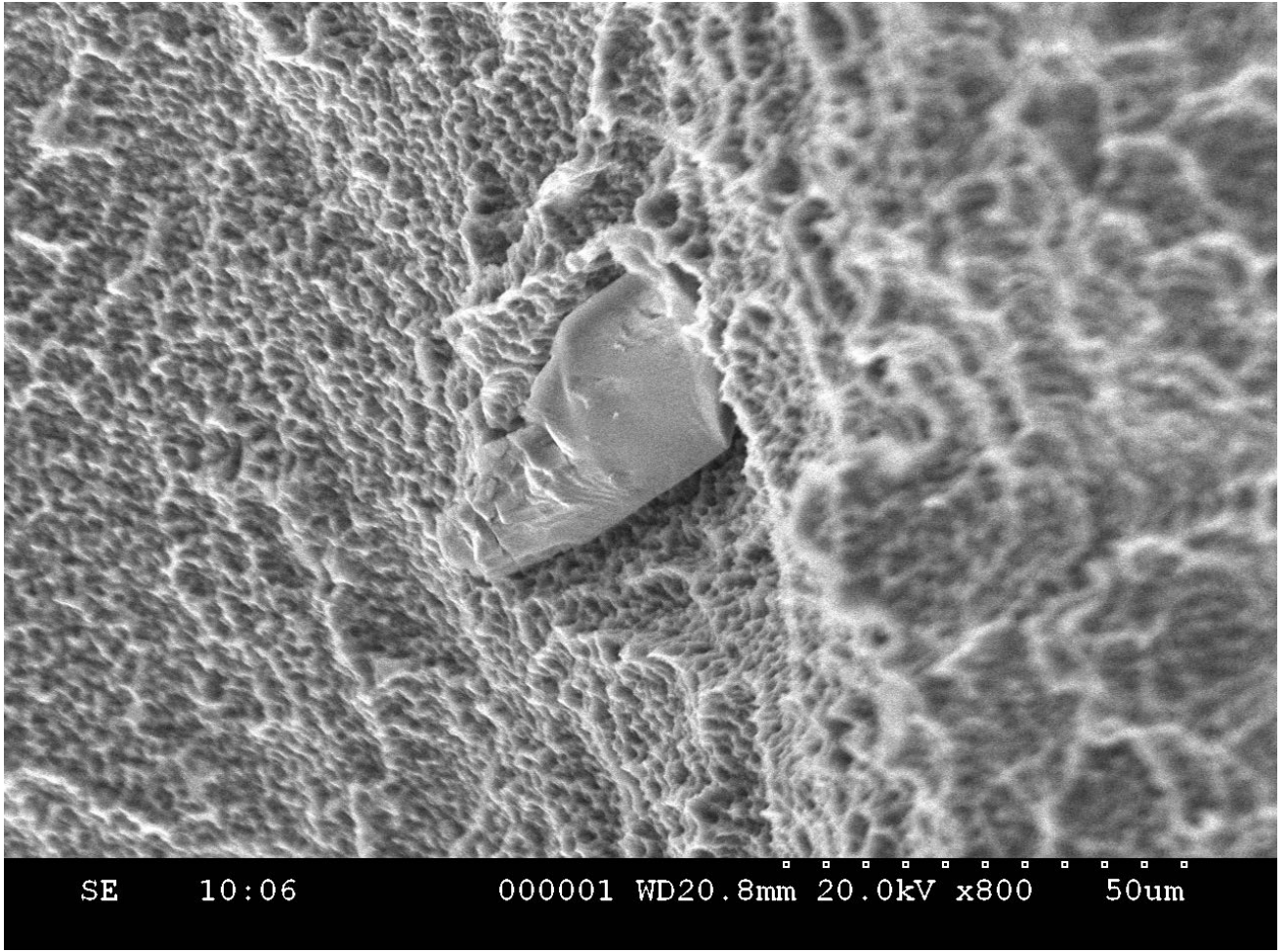

Fig.5: Typical corundum particle struck in the titanium surface without gap (sample #1; 800x).

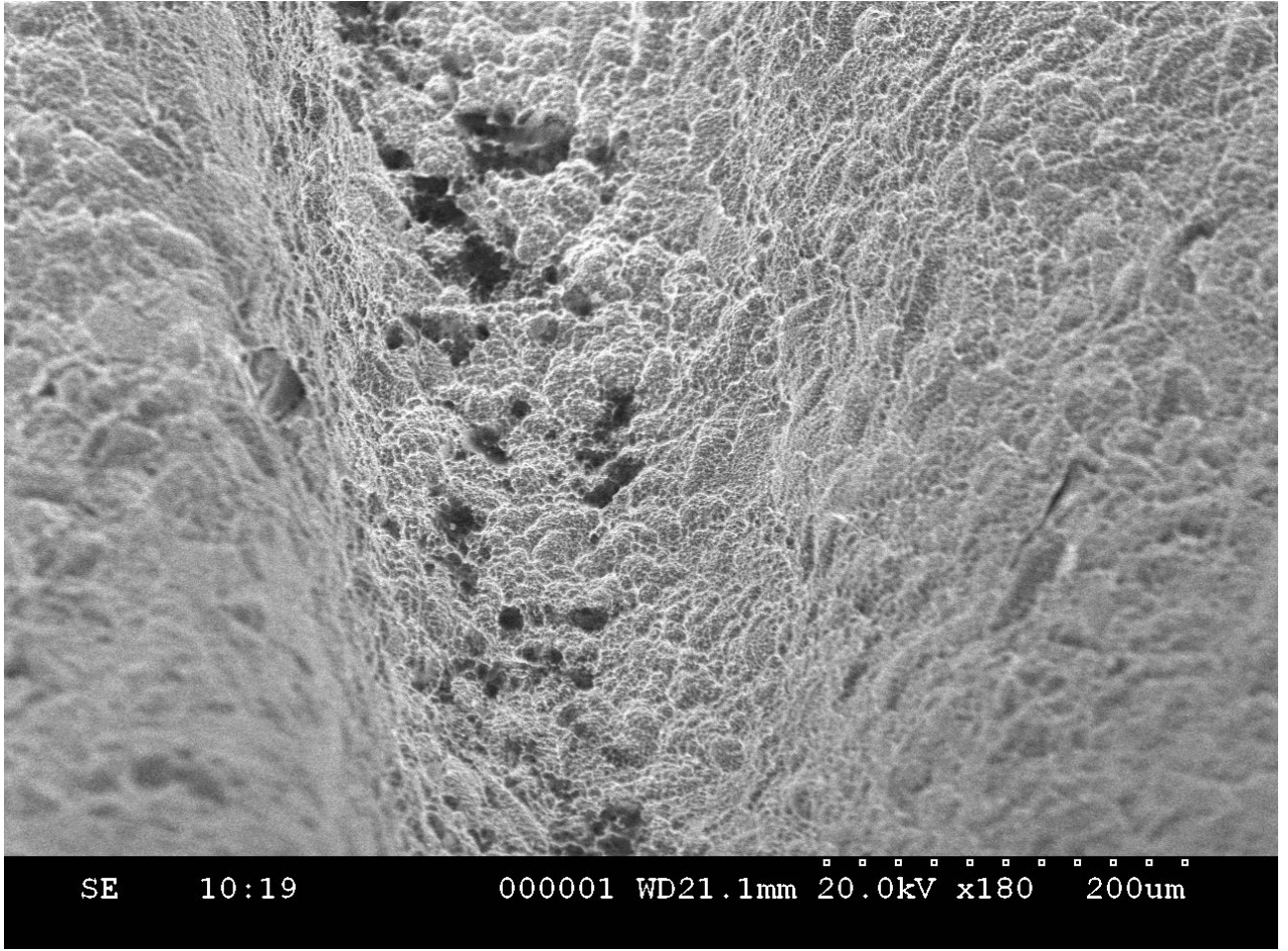

Fig.6: Unusual surface excavations with high carbon levels (sample #1; 180x).

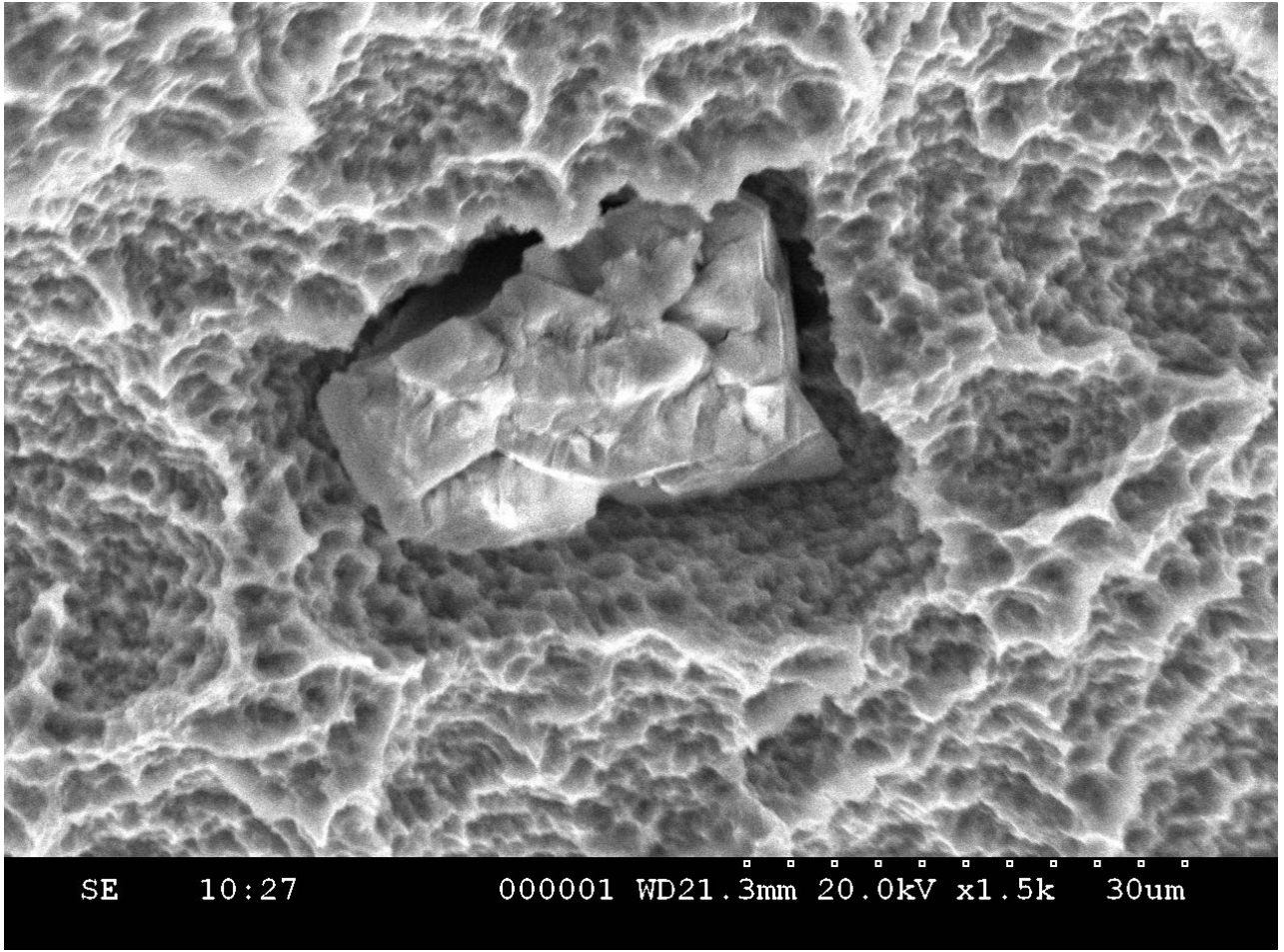

Fig.7: Typical corundum particle struck in the titanium surface with a gap indicating the removal of any material that was located there at the time of blasting (sample #1; 800x).

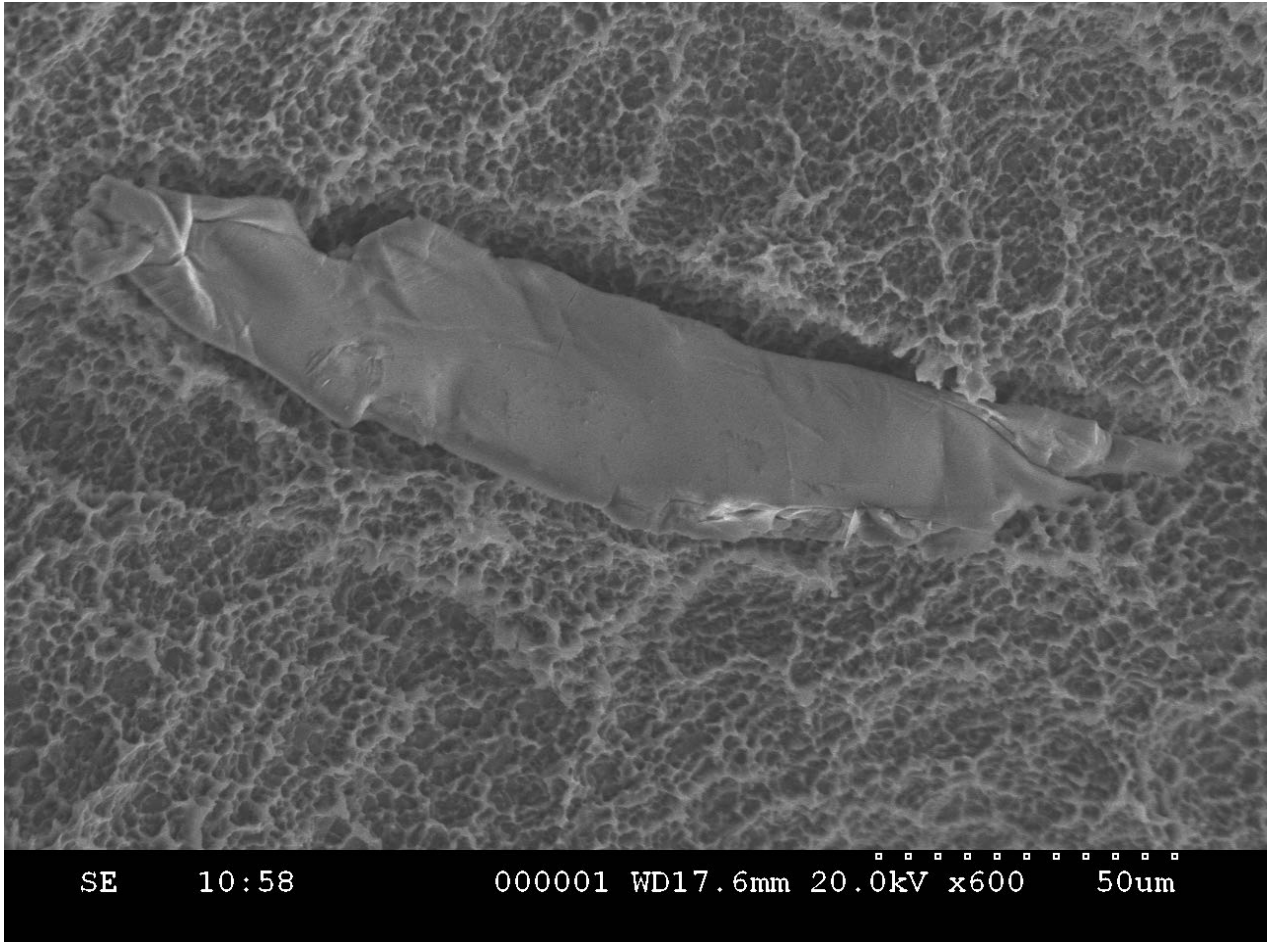

Fig.8: Untypical particle with aluminum content indicating corundum struck in the titanium surface with a gap indicating the removal of any material that was located there at the time of blasting (sample #4; 600x).

## Declaration of interests

☒ The authors declare that they have no known competing financial interests or personal relationships that could have appeared to influence the work reported in this paper.

☐ The authors declare the following financial interests/personal relationships which may be considered as potential competing interests:

Guy Florian Draenert on behalf of all authors

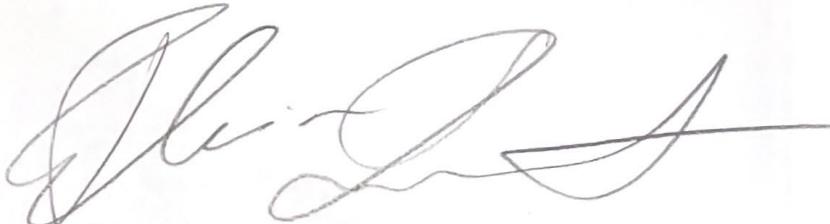

27. my 2022